\documentclass[%
 reprint,
 amsmath,amssymb,
 aps,
prb,
 longbibliography,
 lengthcheck,%
]{revtex4-1}

\usepackage{graphicx}
\usepackage{dcolumn}
\usepackage{bm}
\usepackage{hyperref}

\begin{document}

\preprint{APS/123-QED}

\title{Kinematic formation of the pseudogap
spectral properties in a spatially homogeneous
strongly correlated electron system}

\author{Valery V. Val'kov$^{1,2}$}
\author{Alexander A. Golovnya$^{1,3}$}
\author{Maxim M. Korovushkin$^{1,2}$}%
\affiliation{%
 $^1$L.\,V. Kirensky Institute of Physics, 660036 Krasnoyarsk, Russia\\
 $^2$Siberian Aerospace State University, 660014 Krasnoyarsk, Russia\\
 $^3$Siberian Federal University, 660074 Krasnoyarsk, Russia}%

\date{\today}

\begin{abstract}
It is shown that the kinematic interaction caused by the
quasi-Fermi character of commutation relations for operators of
the atomic representation can induce pseudogap behavior of the
spectral characteristics of an ensemble of Hubbard fermions.
Mathematically, the presence of the kinematic interaction
manifests itself in modification of the faithful representation of
a single-particle Green's function of Hubbard fermions
$D(\textbf{k},i\omega_n)$, which involves, apart from self-energy
operator $\Sigma_L(\textbf{k},i\omega_n)$, strength operator
$P(\textbf{k},i\omega_n)$. It is important that the strength operator
enters both the numerator and the denominator of the exact
expression for $D(\textbf{k},i\omega_n)$. The kinematic
interaction, therefore, not only renormalizes the spectrum of
elementary excitations but significantly affects their spectral
weight. It results in strong modulation of spectral intensity
$A(\textbf{k},\omega)$ occurring on a Fermi contour. Calculations
of the spectral properties for the $t-J$ model in the one-loop
approximation yield good quantitative agreement with the ARPES
data obtained on cuprate superconductors.

\begin{description}
\item[PACS numbers]
71.10.Fd, 71.18.+y, 71.27.+a, 74.40.-n, 74.72.Kf
\end{description}
\end{abstract}

\pacs{71.10.Fd, 71.18.+y, 71.27.+a, 74.40.-n, 74.72.Kf}
\maketitle


\section{\label{sec:level1}Introduction}

The normal phase of cuprate superconductors is characterized by a
number of intriguing properties that cannot be described within
the traditional Fermi liquid theory. These are, first of all, the
pseudogap behavior in an undoped
region~\cite{Timusk99,Sadovskii01,Damascelli03}. It still has been
unclear whether this behavior is explained basing on the modified
Fermi liquid concept or requires building the ground state of a
new type~\cite{Chakravarty01,Simon02}. In the phase diagram of
cuprates, the region of parameters corresponding to the pseudogap
state borders on the region of the superconducting state. This
fact is often interpreted as resemblance of the formation of a
pseudogap and of Cooper instability. Since in many studies the
nature of coupling in cuprate superconductors is attributed to
spin fluctuation processes, these processes are considered to be
the fundamental cause of the formation of the pseudogap state. In
recent years, study of the interrelation between electron and spin
subsystems has become especially important for understanding the
formation of the pseudogap state and establishing the effect of
pseudogap peculiarities of the spectral properties on Cooper
instability. This has been a subject of numerous experimental and
theoretical works on physics of the strongly correlated systems.

The phenomenological concept of the formation of the pseudogap
state proposed in study~\cite{Yang06} uses the idea of the
formation of a quantum spin liquid following the scenario of the
resonance valence bond (RVB) method developed by
Anderson~\cite{Anderson87}. This approach allowed reproducing the
formation of a pseudogap, growing with decreasing doping level,
near the antiferromagnetic Brillouin zone. This process is
accompanied by rearrangement of the large Fermi surface
experimentally observed in the optimal doping region to the small
Fermi or Luttinger pockets arising in the undoped region. This
model was successfully used in the description of the effect of
the pseudogap on certain properties observed in undoped cuprates,
such as electronic specific heat~\cite{LeBlanc09}, London
penetration depth~\cite{Carbotte10}, superconducting
gap~\cite{Schachinger10}, electron density of
states~\cite{Borne10}, electrical and heat
conduction~\cite{Carbotte11}, and anomalies of optical
conduction~\cite{Pound11}. The model was also employed to
interpret the photoemission spectroscopic data
(ARPES)~\cite{Yang09}.

The pseudogap formation was also studied using numerical
calculations on the basis of exact diagonalization~\cite{Maier02},
the quantum Monte Carlo method~\cite{Assaad97,Kyung03}, and the
dynamic mean field theory (DMFT)~\cite{Sadovskii05,Kuchinskii07}.
The results obtained were in satisfactory agreement with the
experimental data, specifically, the presence of a large Fermi
surface at optimal doping~\cite{Stephan91} and modulation of
spectral intensity and reduction of the density of states at the
Fermi level at weak
doping~\cite{Preuss97,Sadovskii05,Kuchinskii07}.

Analysis of the spectral properties of cuprates is often based on
the Hubbard model~\cite{Hubbard63} and its low-energy version, i.
e., the $t-J$ model. Study of the Hubbard model by the
renormalization group method revealed deviation of its spectral
properties from those described within the theory of an ordinary
Fermi liquid and a noticeable decrease in the Fermi surface
area~\cite{Zanchi97,Furukawa98}. The features of the pseudogap
state directly related to antiferromagnetic spin fluctuations
(SFs) were considered in studies~\cite{Schmalian98,Chubukov97}
within the phenomenological spin-fermion model~\cite{Monthoux91}
and by phenomenological investigation of the Hubbard
model~\cite{Dahm99}. In study~\cite{Prelovsek01}, using the method
of equations of motion~\cite{Zubarev60} and Mori's technique of
projection operators~\cite{Mori65}, the spectral functions and
Fermi surface were investigated in the framework of the $t-J$
model. It was shown that long-wavelength SFs govern the
low-frequency behavior of a system, leading to truncation of a
large Fermi surface, modulation of spectral intensity, and a
decrease in the density of states at the Fermi level in the weak
doping region. A microscopic theory for the electron spectrum of
the CuO$_2$ plane within the Hubbard model was proposed in
study~\cite{Plakida07}. In this study the Dyson equation for the
single-electron Green's function in terms of the Hubbard operators
was derived and solved self-consistently for the self-energy
evaluated in the noncrossing approximation. Electron scattering on
spin fluctuations induced by the kinematic interaction was
described by a dynamical spin susceptibility with a continuous
spectrum. At low doping, an arc-type Fermi surface and a pseudogap
in the spectral function close to the Brillouin zone boundary was
observed.

In a number of studies on microscopic investigation of the
spin-fluctuation nature of the pseudogap state, the interaction
between electrons and a spin density wave was used as a mechanism
of the spin-electron correlation. The existence of such a wave was
considered to be an a priori specified property and its origin was
not discussed. Meanwhile, by now the existence of the spin density
wave in cuprate superconductors has not been experimentally
confirmed. In view of this, a reasonable question arises
concerning possible implementation of the pseudogap phase at the
interacting electron and spin degrees of freedom but with no use
of the hypothesis of the spin density wave existing in the system.
To answer this question, one should take into consideration the
important feature of strongly correlated systems, which include
high-temperature superconductors. As is known, in the regime of
strong correlations, the adequate description of electron systems
that takes into account Hubbard correlations is based on the
atomic representation~\cite{Hubbard65}. In this case, the
operators employed in the theory do not satisfy the commutation
relations characteristic of the Fermi operators, since commutation
of two basis operators of the atomic representation results in a
basis operator and not in a number
\begin{equation}
[X_f^{pq},X_m^{rs}]_{\pm}=\delta_{fm}(\delta_{qr}X_f^{ps}\pm\delta_{ps}X_f^{rq}).
\end{equation}
Physically, this feature of the commutation (kinematic) relations
between basis operators manifests itself (for instance, in
derivation of equations of motions or calculation of a scattering
amplitude) as an additional interaction arising in a system. This
interaction, with regard to its nature, is named kinematic. The
occurrence of the kinematic interaction in Heisenberg ferromagnets
was mentioned by Dyson [33]. This interaction occurs in them due
to the noncommutative character of the $SU(2)$ algebra of spin
operators. Taking into account the kinematic interaction, Dyson
performed the correct calculation of a two-magnon scattering
amplitude and obtained valid temperature renormalizations for both
the elementary excitation spectrum and the thermodynamic
characteristics [33].

In cuprate superconductors belonging to the Hubbard strongly
correlated systems, the kinematic interaction manifests wider.
Apart from renormalizing the properties of the normal phase, this
interaction can be a mechanism of Cooper
instability~\cite{Zaitsev87}. The kinematic interaction originates
from the fact that, in the regime of strong correlations, the
Hubbard model is adequately described on the basis of the atomic
representation. In this case, the Hubbard operators are basis. For
them, commutation relations are more complex than those for spin
operators, since they include both quasi-spin and quasi-Bose
operators. Commutation of two quasi-Fermi operators results in a
quasi-Bose operator expressed via the quasi-spin operators and
operators reflecting charge fluctuations. Thus, the kinematic
interaction in the systems of interest couples Hubbard fermions
with charge and spin fluctuations.

The dynamic and kinematic interactions of Hubbard fermions lead to
renormalization of their energy spectrum. Therefore, the faithful
representation for the single-particle Green's function of Hubbard
fermions $D(\textbf{k}, i\omega_n)$  involves, apart form self-energy
operator $\Sigma_L(\textbf{k}, i\omega_n)$ caused by the dynamic
interaction of Hubbard fermions, strength operator $P(\textbf{k},
i\omega_n)$ arising due to the kinematic interaction of these
fermions. It is important that $P(\textbf{k}, i\omega_n)$ enters
both the numerator and the denominator of the faithful
representation for the distinguished Green's function. While the
occurrence of the strength operator in the denominator of
$D(\textbf{k}, i\omega_n)$ directly affects renormalization of the
elementary excitation spectrum, the occurrence of the strength
operator in the numerator of $D(\textbf{k}, i\omega_n)$ determines
renormalization of spectral intensity. This is of fundamental
importance for investigation of the spectral characteristics of
strongly correlated systems.

The above-mentioned coupling of the quasi-Fermi operators of the
atomic representation with the quasi-Bose operators implies that,
physically, the kinematic interaction reflects the presence of the
interaction between charge and spin degrees of freedom and between
Fermi and Bose excitations in a strongly correlated electron
system. Therefore, the calculation of contributions of these
interactions to renormalizations of energies of the elementary
excitations and their spectral intensities is reduced to the
calculation of the self-energy $\Sigma_L(\textbf{k}, i\omega_n)$ and
strength $P(\textbf{k}, i\omega_n)$ operators. This is the specific
way of theoretical study of the pseudogap phase in a spatially
homogeneous case. The present study is aimed at solving this
problem within the $t-J$ model on the basis of the diagram
technique for Hubbard operators~\cite{Zaitsev7576,Zaitsev04}. The
key point of the developed theory is that it uses the faithful
representation for the Matsubara Green's function $D(\textbf{k},
i\omega_n)$ via the self-energy $\Sigma_L(\textbf{k}, i\omega_n)$ and
strength $P(\textbf{k}, i\omega_n)$ operators.

In Section~\ref{sec:level2}, using the modified Dyson equation for
a strongly correlated system, the correlation between the spectral
intensity and the strength and self-energy operators for the $t-J$ model is
established. Then, using the diagram technique for Hubbard
operators, contributions to the strength and self-energy operators are
calculated in the one-loop approximation and the integral equation
for the correction to the strength operator is written. In
Section~\ref{sec:level3}, the choice of magnetic susceptibility
entering the integral equation kernels is discussed. In
Section\ref{sec:level4}, the results of the calculation of the
Fermi excitation spectrum, spectral intensity, Fermi surface, and
density of electron states demonstrating the pseudogap behavior of
the system are reported. Section~\ref{sec:level5} presents the
data for the limiting case of strong electron correlations, which
allows obtaining relatively simple analytical expressions for the
energy of Fermi excitations and spectral intensity. The kinematic
mechanism of the pseudogap state formation is demonstrated in the
microscopic scale. The final section contains discussion of the
results.

\section{\label{sec:level2}Correlation between the spectral
intensity and the strength and self-energy operators}

The Hamiltonian of the $t-J$ model in the atomic representation is
\begin{eqnarray}
\label{Ham} &&\hat H=\hat H_{t}+\hat H_{J},\\
&&\hat H_{t}=\sum_{f\sigma}(\varepsilon-\mu)X^{\sigma\sigma}_f +
\sum_{fm\sigma}t_{fm}X^{\sigma 0}_f X^{0\sigma}_m,\nonumber\\
&&\hat
H_{J}=\frac{1}{2}\sum_{fm\sigma}J_{fm}\left(X^{\sigma\bar{\sigma}}_f
X^{\bar{\sigma}\sigma}_m-X^{\sigma\sigma}_f
X^{\bar{\sigma}\bar{\sigma}}_m\right),\nonumber
\end{eqnarray}
where $X^{pq}_f=|f,p\rangle\langle f,q|$  are the Hubbard
operators~\cite{Hubbard63} describing the transition of an ion in
the  $f$-th site from the one-site state $|f,q\rangle$ to the
state $|f,p\rangle$, $\varepsilon$ is the energy of one-electron
one-ion state, $\mu$ is the chemical potential of the system,
$\sigma=\pm 1/2$ ($\bar{\sigma}=-\sigma $) is the spin moment
projection, $t_{fm}$ is the integral of electron hopping from the
$m$-th to $f$-th site, $J_{fm}=2t_{fm}^2/U$ is the exchange
integral, and $U$ is the Hubbard repulsion parameter.

We calculate spectral intensity $A(\textbf{k},\omega)$ with the
use of the diagram technique for Hubbard
operators~\cite{Zaitsev7576,Zaitsev04}, introducing the Matsubara
Green's function
\begin{eqnarray}\label{Ddefinition}
&&D_{0\sigma,0\sigma}(f,\tau;f',\tau')=-\langle
T_{\tau}\tilde{X}^{0\sigma}_f(\tau)\tilde{X}^{\sigma0}_{f'}(\tau')\rangle=\\
&&=\frac
TN\sum_{\textbf{k}\omega_n}e^{i\textbf{k}(f-f')-i\omega_n(\tau-\tau')}
D_{0\sigma,0\sigma}(\textbf {k},i\omega_n)\nonumber.
\end{eqnarray}
Here $T_{\tau}$ is the operator of Matsubara time ordering. In
Expression (\ref{Ddefinition}), the Hubbard operators are taken in
the Heisenberg representation with Matsubara time $\tau$
\begin{equation}
\label{DeMatsRepr} \tilde{X}^{0\sigma}_{f}(\tau) = \exp(\tau \hat
H) X^{0\sigma}_{f}\exp(-\tau \hat H ),~ 0 < \tau <\frac1T,
\end{equation}
where $T$ and $\hat H$ are the temperature and Hamiltonian of the
system, respectively.

Below, taking into account that the expression for the Green's
function in the paraphase is independent of spin polarization, we
omit spin indices. An important feature of the introduced
functions is that $D(\textbf {k},i\omega_n)$ decomposes into the
product of the propagator part and the strength
operator~\cite{Zaitsev7576}
\begin{equation}
\label{ConnDG} D(\textbf{k},i\omega_n)=G(\textbf{k},i\omega_n)
P(\textbf{k},i\omega_n).
\end{equation}
Thus, it is easy to obtain the modified Dyson equation for
$G(\textbf{k},i\omega_n)$
\begin{equation}
\label{DysEq}
\includegraphics[width=0.45\textwidth]{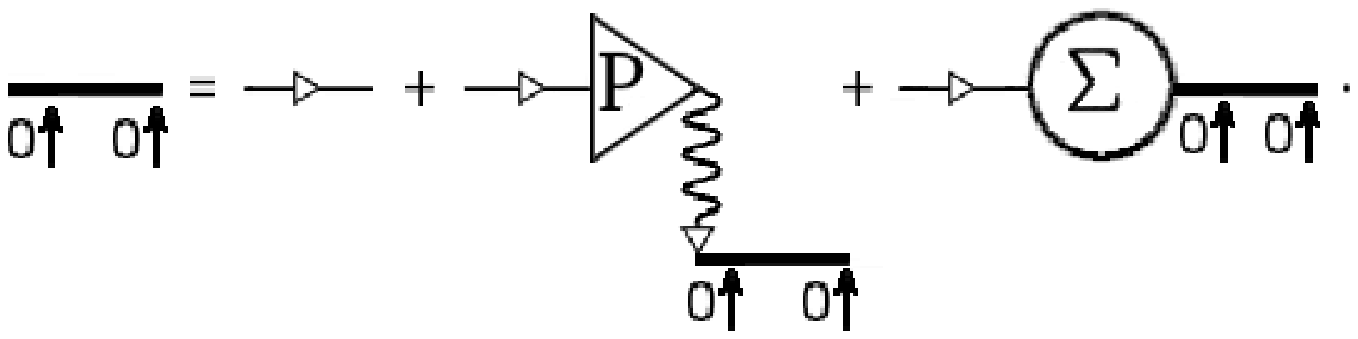}
\end{equation}
The bold line in the equation corresponds to the total propagator
$G(\textbf{k},i\omega_n)$ and the triangle with symbol $P$ denotes
strength operator $P(\textbf{k},i\omega_n)$. The circle with
inscribed symbol $\Sigma^L$ corresponds to the Larkin-irreducible
self-energy operator $\Sigma^L(\textbf{k},i\omega_n)$
~\cite{Baryakhtar84}. The fine line with the light (dark) arrow
denotes the seed Green's function for a Hubbard fermion that
corresponds to the analytical expression
\begin{equation}
G_{0}(i\omega_m)= \frac{1}{i\omega_m-\varepsilon+\mu}.
\end{equation}
The wavy lines with the light and dark arrows denote the Fourier
image of hopping integral $t_{\textbf{k}}$. The total propagator
$G(\textbf{k},i\omega_n)$ relates to the strength and self-energy operators
as~\cite{Zaitsev04,VV01}
\begin{eqnarray}
\label{Relation}
G(\textbf{k},i\omega_n)=\frac{1}{i\omega_m-\xi-t_{\textbf{k}}P(\textbf{k},i\omega_n)-
\Sigma^L(\textbf{k},i\omega_n)},
\end{eqnarray}
where $\xi=\varepsilon-\mu$. Making the analytical continuation
$i\omega_n\rightarrow\omega+i\delta$ and introducing the real and
imaginary parts of the strength and self-energy operators
\begin{eqnarray}
&&P(\textbf{k},i\omega_n)\rightarrow
P(\textbf{k},\omega+i\delta)=P_1(\textbf{k},\omega)+
iP_2(\textbf{k},\omega),\\
&&\Sigma^L(\textbf{k},i\omega_n)\rightarrow\Sigma^L(\textbf{k},\omega+i\delta)=
\Sigma_1^L(\textbf{k},\omega)+i\Sigma_2^L(\textbf{k},\omega),\nonumber
\end{eqnarray}
we arrive at
\begin{eqnarray}\label{D_ret}
D(\textbf{k},\omega+i\delta)=
\frac{P_1(\textbf{k},\omega)+iP_2(\textbf{k},\omega)}
{\omega-\xi-\Sigma_1(\textbf{k},\omega)+i\bigl(\delta-\Sigma_2(\textbf{k},
\omega)\bigr)}.
\end{eqnarray}
In this expression,
\begin{eqnarray}
&&\Sigma_1(\textbf{k},\omega)=t_{\textbf{k}}P_1(\textbf{k},\omega)+\Sigma_1^{L}(\textbf{k},\omega),\nonumber\\
&&\Sigma_2(\textbf{k},\omega)=t_{\textbf{k}}P_2(\textbf{k},\omega)+\Sigma_2^{L}(\textbf{k},\omega),
\end{eqnarray}
the real and imaginary parts of the Dyson-irreducible self-energy
operator, respectively.

Using the representation for the retarded Green's function
(\ref{D_ret}), we find the spectral intensity
\begin{eqnarray} \label{SpInt}
&&A(\textbf{k},\omega)=-\frac {1}{\pi}~\text{Im}~D(\textbf{k},\omega)=\nonumber\\
&&=-\frac {1}{\pi}\left\{
\frac{\left[\omega-\xi-\Sigma_1^{L}(\textbf{k},\omega)\right]\cdot
P_2(\textbf{k},\omega)}
{\left[\omega-\xi-\Sigma_1(\textbf{k},\omega)\right]^2+\left[\delta-
\Sigma_2(\textbf{k},\omega)\right]^2}\right.\nonumber\\
&&-\left.\frac{
\left[\delta-\Sigma_2^{L}(\textbf{k},\omega)\right]\cdot
P_1(\textbf{k},\omega)}
{\left[\omega-\xi-\Sigma_1(\textbf{k},\omega)\right]^2+
\left[\delta-\Sigma_2(\textbf{k},\omega)\right]^2}\right\}.
\end{eqnarray}
This formula establishes the interrelation between the spectral
intensity $A(\textbf{k},\omega)$ and the strength
$P(\textbf{k},\omega)$ and self-energy $\Sigma^L(\textbf{k},\omega)$
operators. Note that the denominator of Expression (\ref{SpInt})
includes the Dyson self-energy operator, whereas the numerator includes
only its Larkin-irreducible part. This resulted from mutual
reduction of the numerator terms that are the product
$P_1(\textbf{k},\omega)\cdot P_2(\textbf{k},\omega)$.
Our calculations demonstrate that the presence of non-zero imaginary part
$P_2(\textbf{k},\omega)$ of the
strength operator in the numerator of expression (\ref{SpInt}) leads to spectral
intensity
dependency of quasi-momentum on the Fermi surface. The pseudogap state of the
normal phase of strongly correlated electron systems is attributed to this
dependence. In the one-loop approximation~\cite{Zaitsev7576,Zaitsev04,VVGA08},
the correction to $P(\textbf{k},i\omega_m)$ caused by the interactions of the
$t-J$ model is determined by the four graphs
\begin{equation}
\label{diagramsP}
\includegraphics[width=0.4\textwidth]{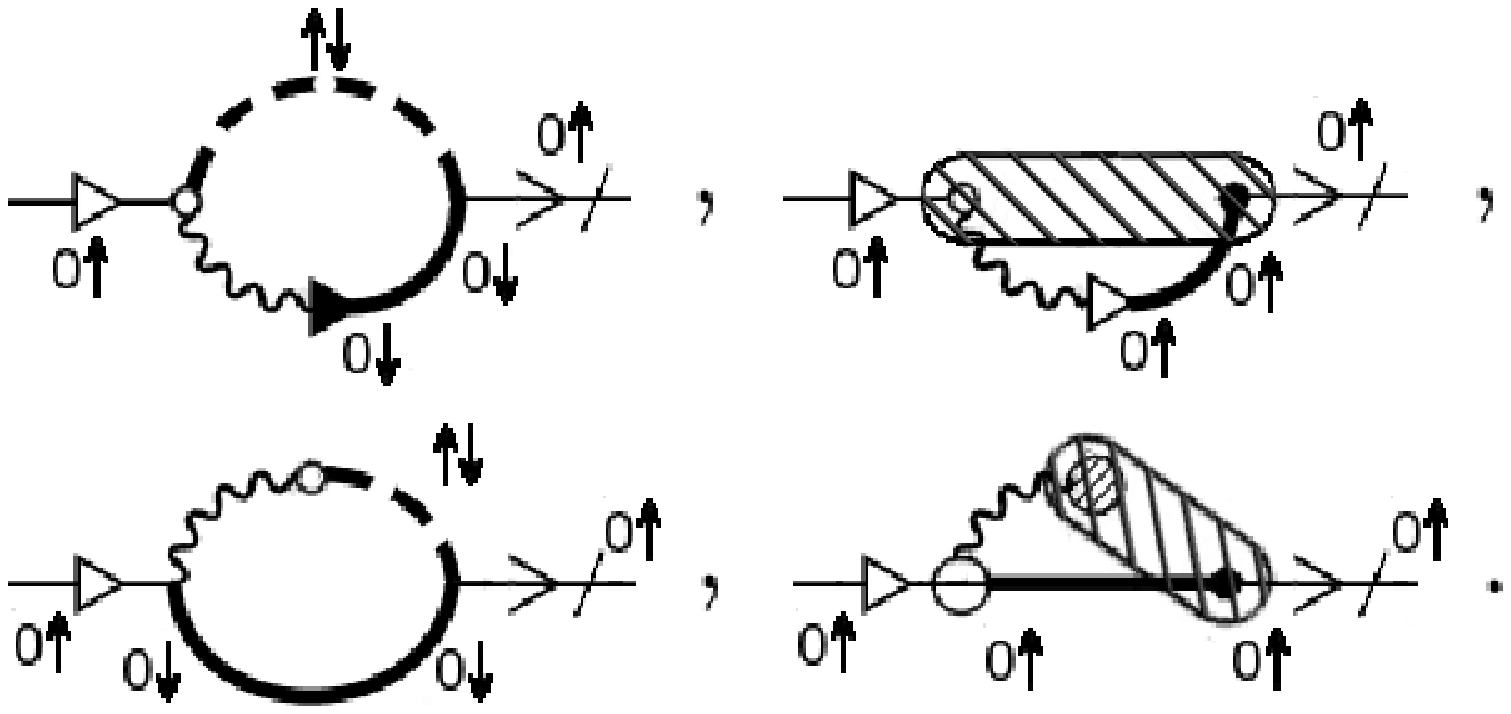}
\end{equation}
The contribution to the self-energy operator $\Sigma^L$ is determined by
the two plots
\begin{equation}
\label{diagramsSigma}
\includegraphics[width=0.4\textwidth]{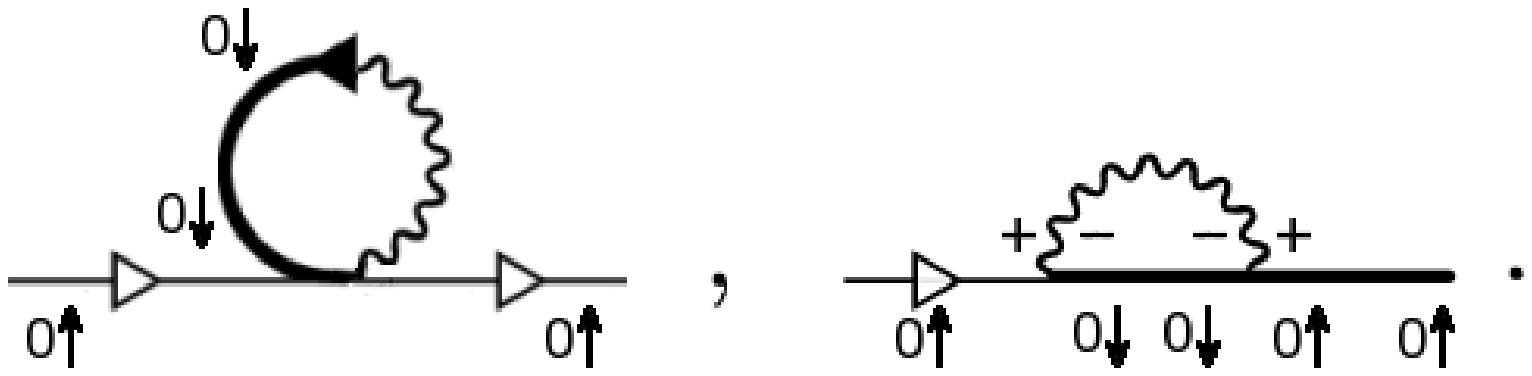}
\end{equation}
In diagrams (13)--(14), the wavy line with the arrow denotes
hopping integral $t_{\bf{q}}$ in the momentum representation. The
end of this line with the dark arrow forms the diagram fragment
induced by the operator $X_f^{0\sigma}$. The wavy lines without
arrows denote the exchange integrals $J_{\bf{q}}$. The
longitudinal interaction
$J_{fm}X_f^{\sigma\sigma}X_m^{\bar{\sigma}\bar{\sigma}}$ is shown
by the wavy line with the two large circles. The end with the
light circle corresponds to the diagram fragment in which the
operator $X_f^{\uparrow\uparrow}$ participated in pairing. The
shaded circle corresponds to the operator
$X_f^{\downarrow\downarrow}$. The transverse interaction
$J_{fm}X_f^{\sigma\bar{\sigma}}X_m^{\bar{\sigma}\sigma}$ is
denoted by the wavy line. At the ends of this line, sequence of
two opposite values of the spin moment projection is shown. This
sequence unambiguously points out that of the two operators
describing the transverse interaction, pairing with which induced
this diagram fragment. The dashed line corresponds to the Fourier
image $D_{\perp}(\textbf{q},i\omega_l)$ of the quasi-spin
transverse Green's function
\begin{eqnarray}
\label{DFSPP} &&-\langle T_{\tau}\tilde{X}_f^{\uparrow
\downarrow}(\tau) \tilde{X}_g^{\downarrow
\uparrow}(\tau')\nonumber
\rangle=\\
&&= \frac{T}{N}\sum_{\textbf{q},\omega_s}
exp\left\{i\left[\textbf{q}(\textbf{R}_f-\textbf{R}_g)-\omega_s(\tau-\tau')\right]
\right\}\nonumber\\
&&\times D_{\perp}(\textbf{q}, i\omega_s),~~\omega_s=2s\pi T,
\end{eqnarray}
and the shaded oval corresponds to the Fourier image of the
Green's function
\begin{equation}
\label{DN0UP} -\left\langle T_{\tau}
\Delta\left(\tilde{X}^{00}_{f}(\tau)+ \tilde{X}^{\uparrow \uparrow
}_{f}(\tau)\right) \Delta\left(\tilde{X}^{00}_{g}(\tau')+
\tilde{X}^{\uparrow \uparrow }_{g}(\tau')\right)\right\rangle.
\nonumber
\end{equation}
In this expression, we used the notation
$$\Delta(\tilde{A}_{f}(\tau))=\tilde{A}_{f}(\tau)-
\langle \tilde{A}_{f}(\tau)\rangle.$$ It can be easily seen that
the function introduced for the oval can be expressed via the spin
longitudinal Green's function and the charge Green's function.
These two functions and their Fourier images are related as
\begin{eqnarray}
\label{DN0UP} &&-\langle T_{\tau}\tilde{S}_f^{z}(\tau)
\tilde{S}_g^{z}(\tau') \rangle=\nonumber\\
&&= \frac{T}{N}\sum_{\textbf{q},\omega_s}
exp\left\{i\left[\textbf{q}(\textbf{R}_f-\textbf{R}_g)-\omega_s(\tau-\tau')\right]
\right\}\nonumber\\
&&\times D_{\parallel}(\textbf{q}, i\omega_s),~~\omega_s=2s\pi T,
\end{eqnarray}

\begin{eqnarray}
\label{DFGRCARG} &&-\left\langle T_{\tau}
\Delta\tilde{n}_{f}(\tau)
\Delta\tilde{n}_{g}(\tau')\right\rangle\nonumber\\
&&= \frac{T}{N}\sum_{\textbf{q},\omega_s}
exp\left\{i\left[\textbf{q}(\textbf{R}_f-\textbf{R}_g)-\omega_s(\tau-\tau')\right]
\right\}\nonumber\\
&&\times C(\textbf{q}, i\omega_s),~~\omega_s=2s\pi T,
\end{eqnarray}
where $$\Delta
n_{f}=X^{\uparrow\uparrow}+X^{\downarrow\downarrow}- \left\langle
X^{\uparrow\uparrow}+X^{\downarrow\downarrow} \right\rangle,$$
$$S^z_f=(X^{\uparrow\uparrow}-X^{\downarrow\downarrow})/2.$$
Associating analytical expressions to plots
(\ref{diagramsP})--(\ref{diagramsSigma}), we obtain the strength and
self-energy operators in the explicit form
\begin{eqnarray}
\label{Expr_P} &&P(\textbf{k},i\omega_m)=C_n+
\frac{T}{N}\sum_{\textbf{q},i\omega_l}(t_{\textbf{q}}+J_{\textbf{k-q}})
G\left(\textbf{q },i\omega_l\right)\times\nonumber\\
&&\times\chi\left(\textbf{q}-\textbf{k}, i\omega_l
-i\omega_m\right),
\\
\label{Expr_Sigma}
&&\Sigma^L(\textbf{k})=-\frac{T}{N}\sum_{\textbf{q},i\omega_l}(t_{\textbf{q}}+J_
{\textbf{k-q}}) G\left(\textbf{q},i\omega_l\right),
\end{eqnarray}
where $C_n=1-n/2$ is the Hubbard renormalization.

These expressions and the faithful representation of the Green's
function $G\left(\textbf{q },i\omega_l\right)$ show that to solve
the equation determining the strength operator, one should know the
spin-charge susceptibility
\begin{equation}
\chi(\textbf{q},i\omega_m)=\chi_{SF}(\textbf{q},i\omega_m)+\chi_{CF}(\textbf{q},
i\omega_m),
\end{equation}
that determines the contribution of the fluctuation processes. For
convenience, this expression contains the dynamic spin
susceptibility
\begin{eqnarray}\label{chiSF}
\chi_{SF}(\textbf{q},i\omega_m)&=&-D_{\perp}(\textbf{q},i\omega_m)-
D_{\parallel}(\textbf{q},i\omega_m)\nonumber\\
&=&-3D_{\parallel}(\textbf{q},i\omega_m),
\end{eqnarray}
and the dynamic charge susceptibility
\begin{equation}
\chi_{CF}(\textbf{q},i\omega_m)=\frac14C(\textbf{q},i\omega_m).
\end{equation}
Expression (\ref{chiSF}) was written taking into account the
equality
$D_{\perp}(\textbf{q},i\omega_m)=2D_{\parallel}(\textbf{q},i\omega_m)$,
since without magnetic field and the long-range magnetic order,
the Hamiltonian of the system is invariant relative to the
transformation of the $SU(2)$ group~\cite{Barabanov92,Plakida99}.

Below, taking into account that the energy of charge excitations
is relatively large, we limit our consideration to the
contributions related to SFs. In this case, the nonlinear integral
equation for the one-loop correction to the strength operator $\delta
P(\textbf{k},i\omega_m)=P(\textbf{k},i\omega_m)-C_n$ is
\begin{eqnarray}
&&\delta P(\textbf{k},i\omega_m)=\nonumber\\
&&=\frac{T}{N}\sum_{\textbf{q},i\omega_l} \frac{(t_{\textbf{q}}+J_
{\textbf{k-q}})\chi_{SF}(\textbf{q}-\textbf{k},
i\omega_l-i\omega_m)}{
i\omega_l-\xi_{\textbf{q}}-t_{\textbf{q}}\,\delta
P(\textbf{q},i\omega_m)-\Sigma^L(\textbf{k})}, \label{dP}
\end{eqnarray}
where $\xi_{\textbf{q}}=\varepsilon+C_nt_{\textbf{q}}-\mu$ and the
Fourier image of the hopping integral is
\begin{eqnarray}
&&t_{\textbf{q}}=t_{1\textbf{q}}+R_{\textbf{q}},~~~
t_{1\textbf{q}}=2 t (\textrm{cos} \,q_x +\textrm{cos} \,q_x),\nonumber\\
&&R_{\textbf{q}}=4 t' \textrm{cos} \,q_x \textrm{cos} \,q_y+2 t''
(\textrm{cos} \,2q_x +\textrm{cos} \,2q_y).
\end{eqnarray}

\section{\label{sec:level3}Magnetic susceptibility}

Since the kernel of the integral equation is determined by the
dynamic magnetic susceptibility, let us briefly analyze this
function. Susceptibility $\chi_{SF}(\textbf{q},i\omega_m)$ for the
Hubbard model was calculated first in~\cite{Hubbard68}. Later,
during the intense studies of high-temperature superconductivity,
$\chi_{SF}(\textbf{q},i\omega_m)$ was calculated in
\cite{Shimahara92, Zaitsev04, Vladimirov07,Eremin08}. The results
of these studies show that the function
$\chi_{SF}(\textbf{q},i\omega_m)$ rapidly drops with increasing
Matsubara frequency. Therefore, the main contribution to the
integral equation is collected by summation over $\omega_l$ close
in value to $\omega_m$. Hence, we may assume that
\begin{eqnarray}
\label{Chi_dyn_sf}
\chi_{SF}\left(\textbf{q},i\omega_m\right)=\chi(\textbf{q})\bar{n}_{SF}
\cdot\delta_{m0}.
\end{eqnarray}
The expression for $\chi(\textbf{q})$ is the spin susceptibility
at zero Matsubara frequency $\bar{n}_{SF}\sim \Omega_{SF}/T$,
where $\Omega_{SF}$ is the value of the Matsubara frequency
starting from which the susceptibility rapidly drops. The order of
magnitude of this frequency is determined by the characteristic
values of excitation energies in a spin subsystem
$\Omega_{SF}\sim0.01|t|$.

\begin{figure}[ht]
\begin{center}
\includegraphics[width=0.35\textwidth]{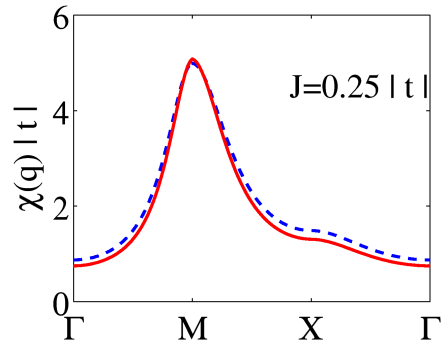}
\caption{(Color online) Quasimomentum dependences of spin
susceptibilities: calculation by the $t-J$ model according to
\cite{Vladimirov07} (dashed line) and calculation by model
susceptibility (\ref{chi}) (solid line). The chosen direction for
rounding the Brillouin zone is $\Gamma(0,0)\rightarrow
M(\pi,\pi)\rightarrow X(\pi,0)\rightarrow\Gamma(0,0)$.}
\label{fig_chi}
\end{center}
\end{figure}

It is important for further consideration that, in the weak doping
region, the $\chi(\textbf{q})$ dependence in the $t-J$ model is
characterized by a sharp peak in the vicinity of the
antiferromagnetic instability point $\textbf{Q}=(\pi,\pi)$.
Results of the numerical calculations of the $\chi(\textbf{q})$
dependence using the technique from~\cite{Vladimirov07} are shown
by the dashed line in Fig.~\ref{fig_chi}. They are in good
agreement with the experimental data~\cite{Hammel89}.

To accelerate the numerical calculation in solving integral
equation (\ref{dP}), we used the model
susceptibility~\cite{Jaklic95,Plakida03}:
\begin{eqnarray}
\label{chi}
&&\chi(\textbf{q})=\frac{\chi_{0}(\xi)}{1+\xi^2(1+\gamma_{1\textbf{q}})},
\end{eqnarray}
where
\begin{eqnarray}\label{chi0}
&&\chi_{0}(\xi)=\frac{3n}{4\omega_s
C(\xi)},~~~ \omega_s=0.55|t|,~~~\xi=2.7,\\
&&C(\xi) = \frac{1}{N}\sum_{\textbf{q}}\frac{1}{1+\xi^2
(1+\gamma_{1\textbf{q}})},~\gamma_{1\textbf{q}}=\frac{(\cos q_x +
\cos q_y)}{2}.\nonumber
\end{eqnarray}
Validity of this approximation follows from comparison of the
dashed and solid lines in Fig.~\ref{fig_chi}.

With allowance for the above assumptions, we obtain, in the first
Born approximation, that
\begin{equation}
\label{delta_Psf}\delta
P(\textbf{k},i\omega_m)=\frac{1}{N}\sum_{\textbf{q}}\frac{(t_{\textbf{q}}+J_
{\textbf{k-q}})\chi(\textbf{q}
-\textbf{k})\Omega_{SF}}{i\omega_m-\xi_{\textbf{q}}},
\end{equation}
\begin{equation}
\label{Sigma_k}
\Sigma(\textbf{k})=-\frac{1}{N}\sum_{\textbf{q}}(t_{\textbf{q}}+J_
{\textbf{k-q}}) n_F(\xi_{\textbf{q}}),
\end{equation}
where $n_F(x)=(\exp(\frac{x-\mu}{T})+1)^{-1}$ is the Fermi--Dirac
function. Then, expression (\ref{ConnDG}) for the single-particle
Green's function acquires the form
\begin{eqnarray}
\label{D_sf} &&D(\textbf{k},i\omega_m)=\frac{C_n+\delta
P(\textbf{k},i\omega_m)}{i\omega_m
  -\xi_{\textbf{k}}-t_{\textbf{k}} \delta
P(\textbf{k},i\omega_m)-\Sigma(\textbf{k})}.
\end{eqnarray}
The obtained system should be added with the equation for chemical
potential $\mu$
\begin{eqnarray}
&&\frac{n}{2}=\frac{T}{N}\sum_{k,\omega_m}
e^{i\omega_m\delta}D(\textbf{k},i\omega_m),~~~\delta\rightarrow\infty.
\end{eqnarray}
The spectral intensity of the system can be calculated after
analytical continuation using expression (\ref{SpInt}).

\section{\label{sec:level4}Pseudogap behavior of the spectral intensity of Hubbard fermions}

\begin{figure}[ht]
\begin{center}
\includegraphics[width=0.45\textwidth]{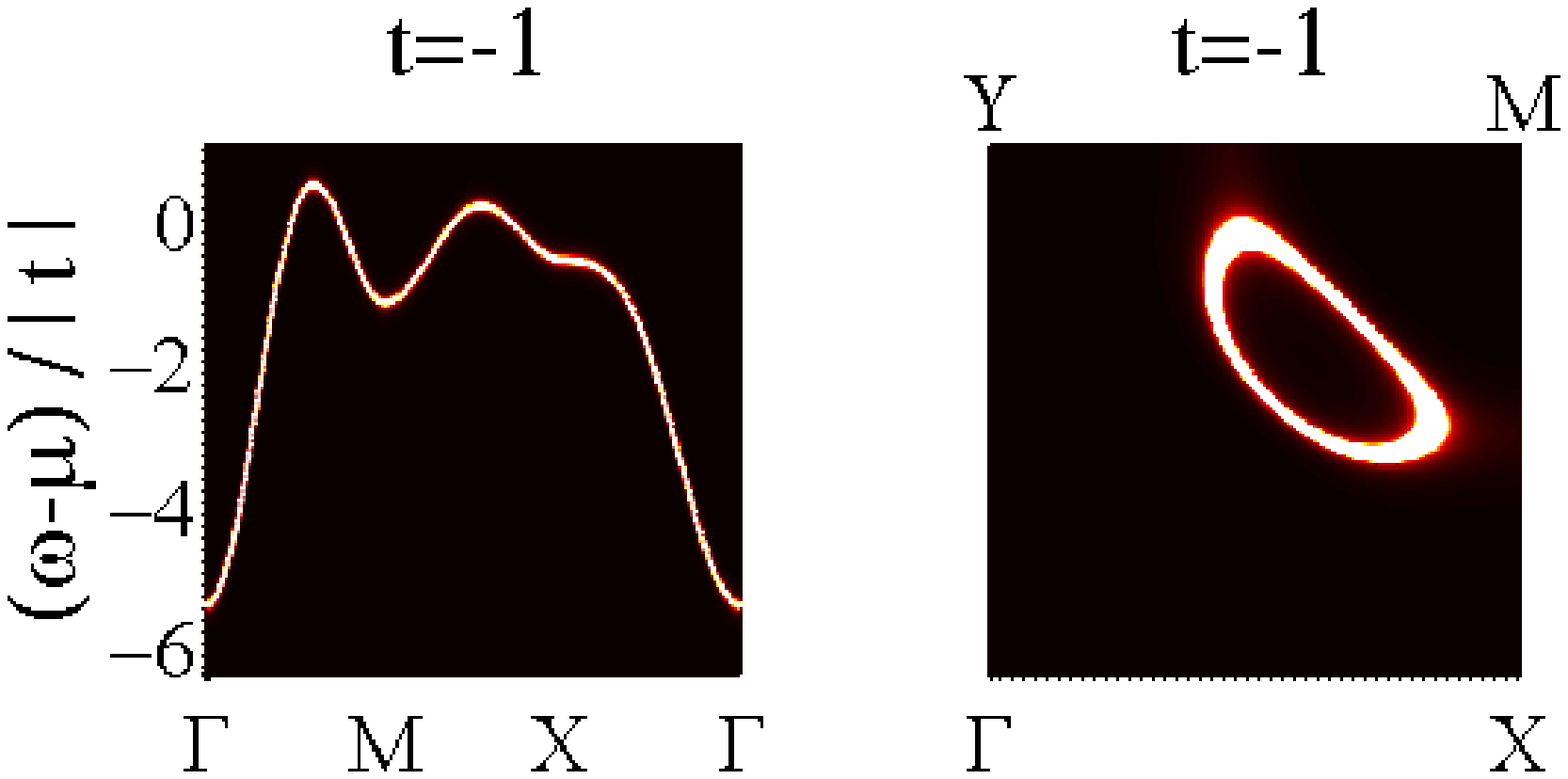}
\includegraphics[width=0.45\textwidth]{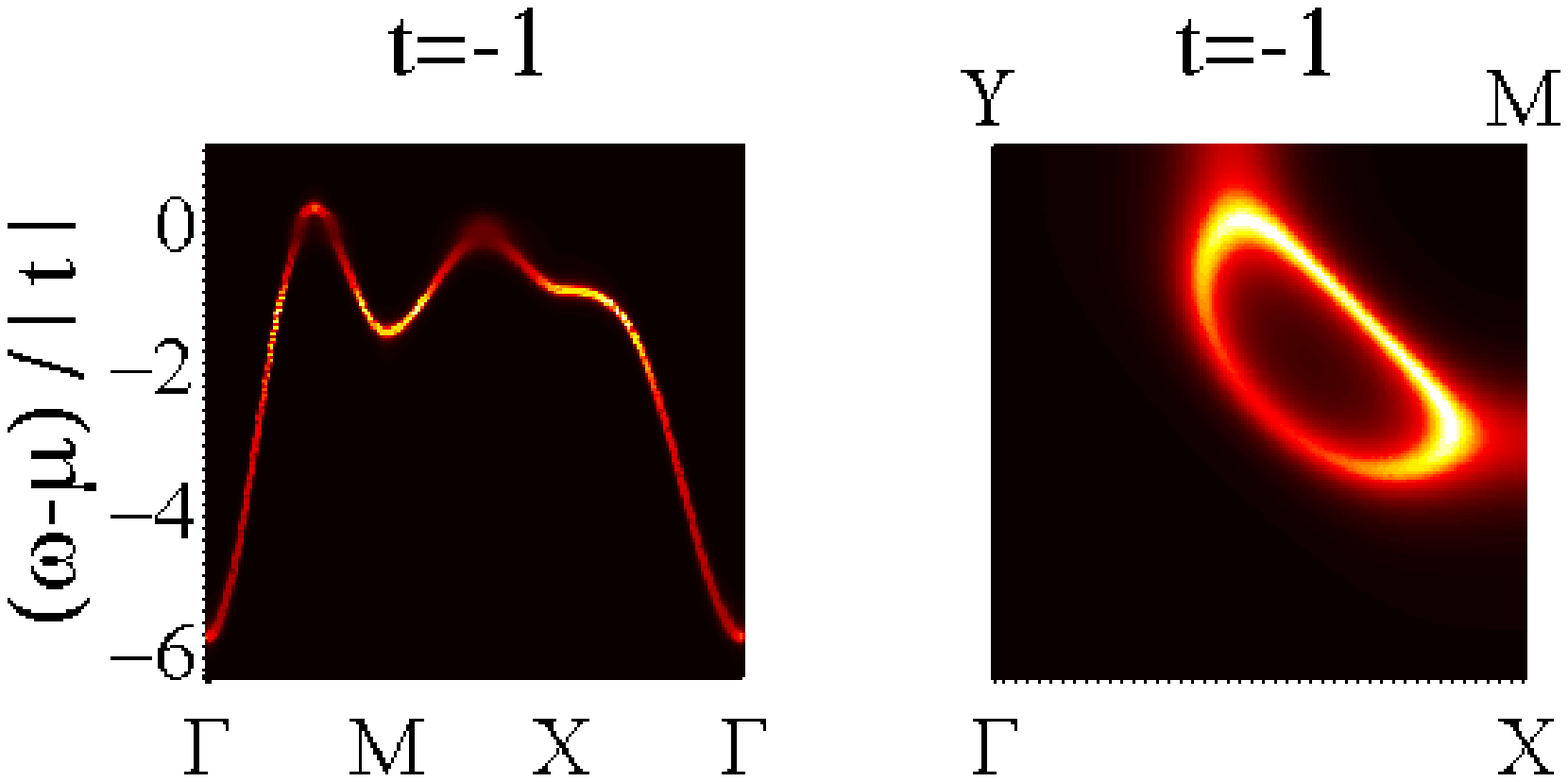}
\includegraphics[width=0.45\textwidth]{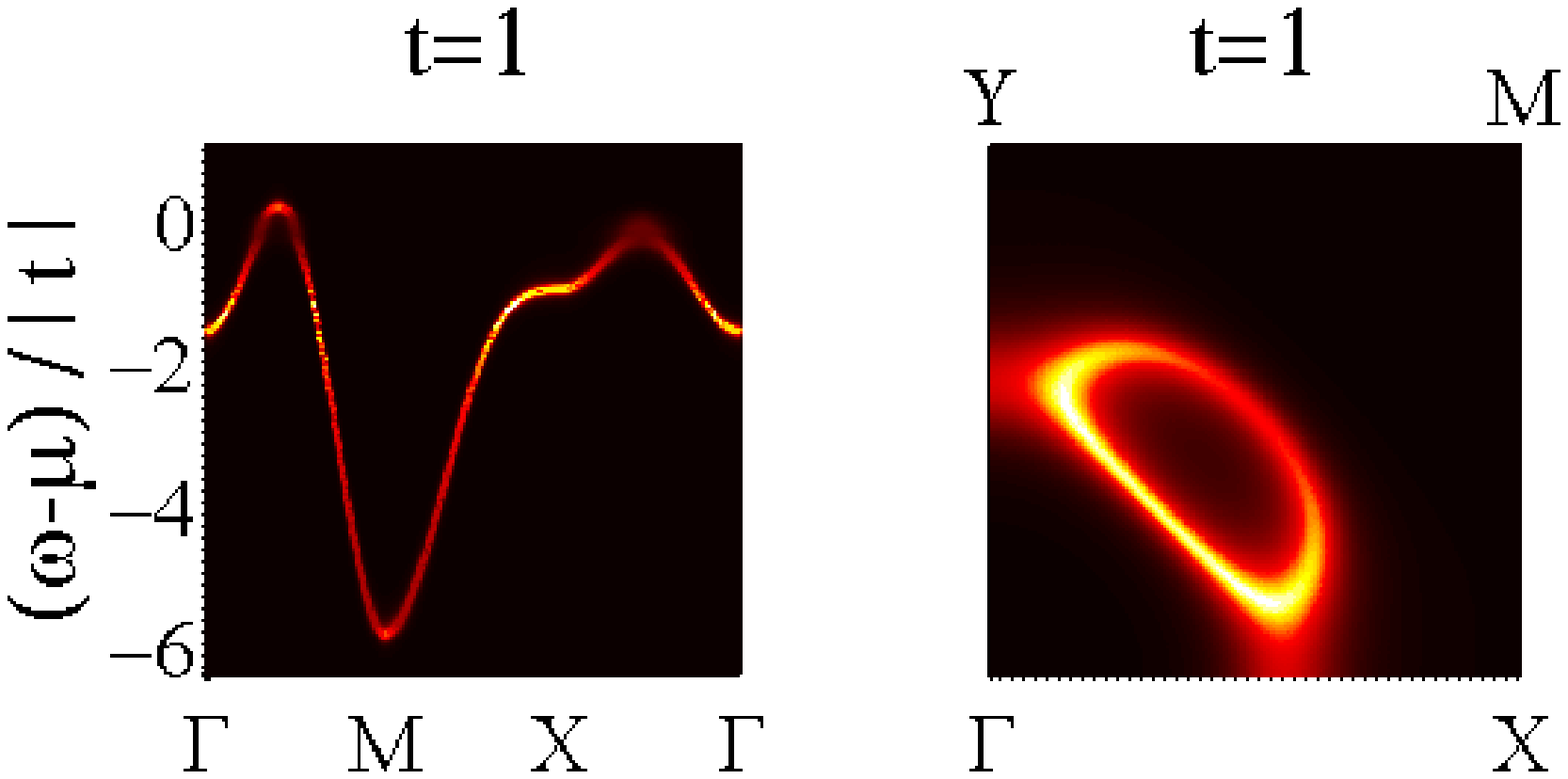}
\caption{(Color online) Lines of the Fermi excitations and Fermi
surfaces with allowance for the spectral intensity. The upper
panel corresponds to the Hubbard-I approximation; the middle and
lower panels, to allowance for SFs at different signs of hopping
parameter $t$.} \label{SI_FS_SF}
\end{center}
\end{figure}

Fig. \ref{SI_FS_SF} presents spectral intensities
$A(\textbf{k},\omega)$ of the model under consideration calculated
for the electron density $n=0.95$ along the principle directions
of the Brillouin zone (left plots) and on the Fermi contour for a
quarter of the Brilouin zone (right plots). Here, the value of the
spectral intensity is reflected by brightness of the energy
spectrum lines (the lighter the line, the larger is the value
$A(\textbf{k},\omega)$). The energy parameters of the model was
measured in units $|t|$
\begin{eqnarray}\label{Data}
T=0.01,~t=-1,~t'=-0.65,~t''=-0.5,~J=0.25\nonumber
\end{eqnarray}
and chosen such that the Fermi surface would have the form of
pockets, in accordance with the experimental data on magnetic
oscillations~\cite{Doiron07}. The upper panel shows
$A(\textbf{k},\omega)$ corresponding to the Hubbard-I
approximation. In this case, the value of $A(\textbf{k},\omega)$
remains invariable at the change in the quasimomentum along the
energy spectrum and at the Fermi surface. The middle panel
demonstrates the results of the calculation of
$A(\textbf{k},\omega)$ with regard to SFs. Comparison with the
upper panel shows that the allowance for SFs results in the
qualitative difference, specifically, the occurrence of
considerable $A(\textbf{k},\omega)$ modulation both at the
spectrum line and at the Fermi level. It can be seen that the
value of $A(\textbf{k},\omega)$ decreases most in the wide energy
region near the chemical potential.

\begin{figure}[ht]
\begin{center}
\includegraphics[width=0.35\textwidth]{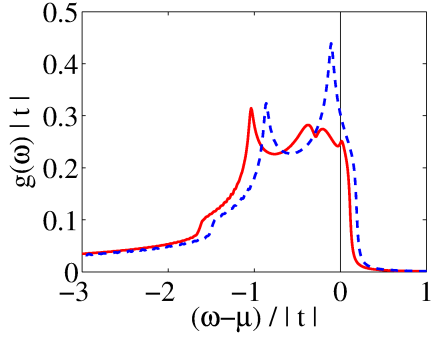}
\caption{(Color online) Density of electron states calculated in
the Hubbard-I approximation (dashed line) and with allowance for
the spin-fluctuation processes (solid line). The vertical line
reflects the position of the chemical potential.} \label{DOS_SF}
\end{center}
\end{figure}

Note an important feature related to the effect of the sign of $t$
on modulation of the spectral intensity. This feature is
illustrated in the lower panel of Fig.~\ref{SI_FS_SF}, where the
spectral intensity calculated for positive $t$ is shown. In the
calculation, all the rest parameters of the system remained
invariable. Comparison of the middle and lower panels of the
figure shows that at negative $t$ the spectral intensity at the
Fermi contour in maximum in the vicinity of the point $(\pi,\pi)$,
whereas at $t>0$ the $A(\textbf{k},\omega)$ maximum at the Fermi
contour is located on the opposite side of the pocket, i.e., at
the contour fragment that is closer to the point $(0,0)$. Note
that this case corresponds to the results of the ARPES experiments
(see, for example,~\cite{Hashimoto08}).

The developed modification of the spectral intensity by the
expense of the fluctuation processes qualitatively changes the
density of electron states
\begin{equation}
\label{SI} g(\omega)=\frac{1}{N}\sum_\textbf{k}
A(\textbf{k},\omega).
\end{equation}
Fig.~\ref{DOS_SF} illustrates the densities of states calculated
in the Hubbard-I approximation (dashed line) and with allowance
for SFs (solid line). Comparing the two curves, one can see that
the allowance for the fluctuation processes results in the drastic
reduction of the density of states in the vicinity of the chemical
potential. Thus, the results of our analysis show the formation of
the pseudogap state in the considered system of Hubbard fermions.

\section{\label{sec:level5}Spectrum of Hubbard fermion excitations
 and the pseudogap behavior in the limit $U \rightarrow \infty$}

In this Section, we thoroughly analyze the physical nature of
modulation of the spectral intensity in the system of Hubbard
fermions. To elucidate the key points, we consider a simplified
problem allowing us to obtain analytical expressions for the Fermi
energy spectrum and spectral intensity but, at the same time,
preserving the fundamental features of Hubbard fermions. It is
possible with the use of the Hubbard model (\ref{Ham}) in the
regime of extremely strong electron correlations ($U \rightarrow
\infty$) when the Hamiltonian contains only the part corresponding
to the operator of kinetic energy in the atomic representation
($t$ model)
\begin{equation}\label{Hamtmodel}
\hat H=\hat H_{t}.
\end{equation}
Note that, since the model is simplified down to the limit, the
below results on the properties of the Fermi excitation spectrum
cannot be applied to the description of the experimental ARPES
data and are used only to explain clearly the physical nature of
the modulation of the spectral intensity in the ensemble of
Hubbard fermions. This approach is analogous, in a sense, to the
way in which the idealized Kronig--Penney model~\cite{Kronig31}
illustrates the nature of the occurrence of energy bands in a
crystal at the analysis of electron motion in a periodic
potential.

To study the spectral properties of Hubbard fermions in the $t$
model, we use, as earlier, the method developed in
Section~\ref{sec:level2}. In our case, the one-loop correction for
$\delta P(\textbf{k},i\omega_m)$ is determined only by the two
upper plots from (\ref{diagramsP}). The corresponding nonlinear
integral equation for the strength operator will have the form
\begin{eqnarray}
\delta
P(\textbf{k},i\omega_m)=\frac{T}{N}\sum_{\textbf{q},i\omega_l}
\frac{t_{\textbf{q}}\,\chi(\textbf{q}-\textbf{k},i\omega_l
-i\omega_m)}{i\omega_l-\xi_{\textbf{q} }-t_{\textbf{q}}\,\delta
P(\textbf{q},i\omega_m)}. \label{dP}
\end{eqnarray}

Choosing the quasimomentum dependence for the spin susceptibility,
we again take into account the experimental fact of the presence
of a sharp peak of this function in the vicinity of the
antiferromagnetic instability point
$\textbf{Q}=(\pi,\pi)$~\cite{Hammel89}. However, unlike the
previous case, now we obtain simplified analytical expressions
considering the limiting situation where the peak of magnetic
susceptibility is delta-like. Then, $\chi_{SF}(\textbf{q},0)$ is
presented as
\begin{equation}
\label{Chi_stat}\chi_{SF}\left(\textbf{q},0\right)=\chi_0\cdot
\delta(\textbf{q}-\textbf{Q}).
\end{equation}
To determine the value of $\chi_0$, we use, as earlier, model
susceptibility (\ref{chi0}). Then, for the first Born
approximation, we have
\begin{equation}
\label{delta_Psf}\delta
P(\textbf{k},i\omega_m)=\frac{\Omega_{SF}\chi_0
t_{\textbf{k}+\textbf{Q}}}{i\omega_m-\xi_{\textbf{k}+\textbf{Q}}}
\end{equation}
and expression (\ref{ConnDG}) for the single-particle Green's
function acquires the two-pole structure
\begin{eqnarray}
\label{D_sf} &&D(\textbf{k},i\omega_m)=\frac{C_n+\delta
P(\textbf{k},i\omega_m)}{i\omega_m
  -\xi_{\textbf{k}}-t_{\textbf{k}} \delta P(\textbf{k},i\omega_m)}=\\
&&=\frac{C_n(i\omega_m-\xi_{\textbf{k}+\textbf{Q}})+\Omega_{SF}\chi_0
t_{\textbf{k}+\textbf{Q}}}{(i\omega_m)^2
  +\xi_{\textbf{k}}
\xi_{\textbf{k}+\textbf{Q}}-i\omega_m(\xi_{\textbf{k}}+\xi_{\textbf{k}+\textbf{Q}})
-\Omega_{SF}\chi_0
t_{\textbf{k}}t_{\textbf{k}+\textbf{Q}}}.\nonumber
\end{eqnarray}
Hence, the allowance for SFs forms two branches of the Fermi
excitation spectrum
\begin{equation}
\label{Spectr_SF} E^{\pm}_{\textbf{k}}=C_nR_{\textbf{k}} \pm
\sqrt{\left(C_n^2-\Omega_{SF}\chi_0 \right)t_{1\textbf{k}}^2
+\Omega_{SF}\chi_0 R^2_{\textbf{k}}}.
\end{equation}
The equation for chemical potential $\mu$ is
\begin{eqnarray}
&&\frac{n}{2}=\frac{1}{N}\sum_k\Bigl(n_F(E^+_{\textbf{k}})A^+_{\textbf{k}}
+n_F(E^-_{\textbf{k}})A^-_{\textbf{k}} \Bigr),
\end{eqnarray}
where the functions
\begin{eqnarray}
&&A^{\pm}_{\textbf{k}}=\frac{1}{2} (C_n\pm\lambda_{\textbf{k}}),\nonumber\\
&&\lambda_{\textbf{k}}=\frac{(C_n^2-\Omega_{SF}\chi_0)t_{1\textbf{k}}+\Omega_{SF}\chi_0
R_{\textbf{k}} } {\sqrt{\left(C_n^2-\Omega_{SF}\chi_0
\right)t_{1\textbf{k}}^2 +\Omega_{SF}\chi_0 R_{\textbf{k}}^2}}
\end{eqnarray}
determine partial contributions of each branch of the spectrum to
the total spectral intensity
\begin{eqnarray*}
&&A(\textbf{k},\omega)=A^{+}(\textbf{k},\omega)+A^{-}(\textbf{k},
\omega) ,\\
&&A^{\pm}(\textbf{k},\omega)=A^{\pm}_{\textbf{k}}\delta(\omega-E^{\pm}_{\textbf{k}}).
\end{eqnarray*}
As will be shown below, the change in the values of the
$A^{\pm}_{\textbf{k}}$ functions at the change in  $\textbf{k}$
(for example, upon motion along the isoenergetic line) leads to
the formation of significant modulation of $A(\textbf{k},\omega)$
as soon as SFs become strong.
\begin{figure}[ht]
\begin{center}
\includegraphics[width=0.44\textwidth]{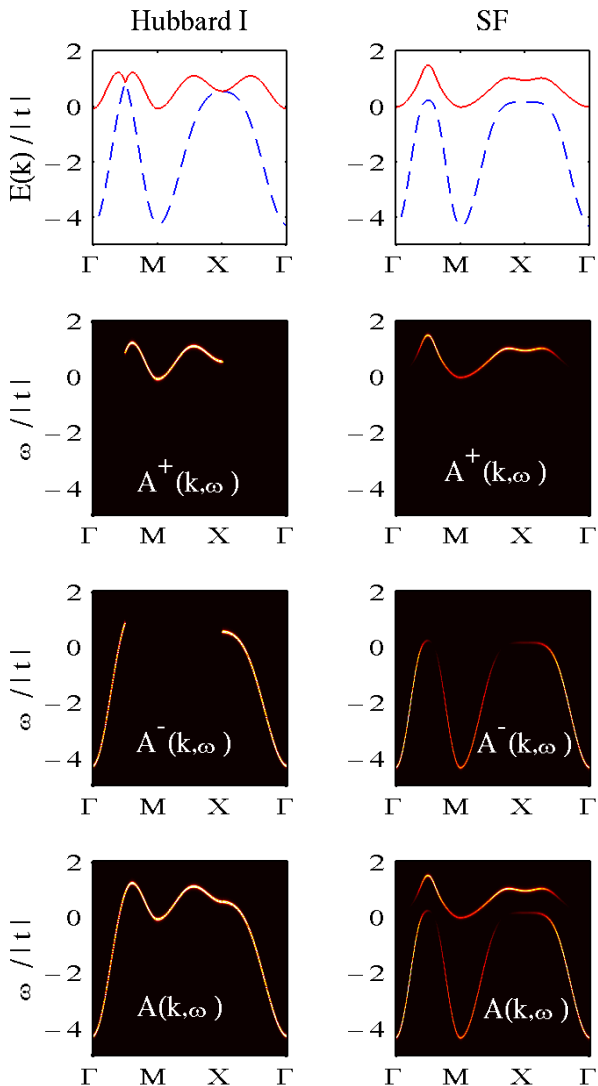}
\caption{(Color online) Spectrum of the Fermi excitations along
the principal directions of the Brillouin zone and modification of
the spectral intensities with allowance for the spin-fluctuation
processes.} \label{fig_spectra_SF}
\end{center}
\end{figure}

\begin{figure}[ht]
\begin{center}
\includegraphics[width=0.35\textwidth]{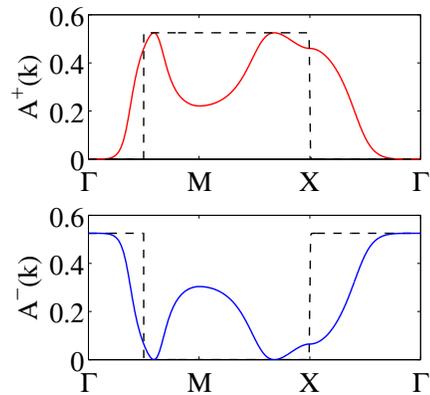}
\caption{(Color online) Dependences of partial contributions to
the spectral intensity on the quasimomentum with allowance for SFs
(solid lines). For comparison, the dashed lines show the
dependences of these functions in the Hubbard-I approximation.}
\label{A_k}
\end{center}
\end{figure}

\begin{figure}[ht]
\begin{center}
\includegraphics[width=0.45\textwidth]{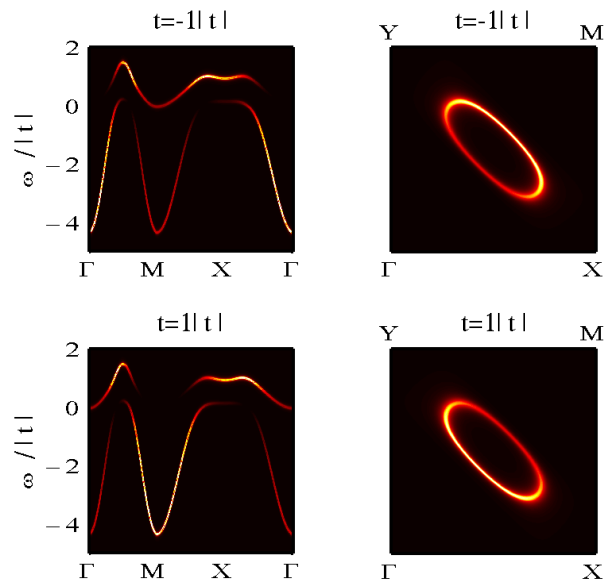}
\caption{(Color online) Effect of the sign of hopping parameter
$t$ on modulation of the spectral intensity.} \label{SI_FS_SF2}
\end{center}
\end{figure}

The occurrence of $A(\textbf{k},\omega)$ modulation is illustrated
in Fig.~\ref{fig_spectra_SF}. The upper plot on the left presents
the energy spectrum development in the limit of the vanishingly
small power of SFs ($\Omega_{SF}\cdot \chi_0\rightarrow 0$) for
the electron density $n=0.95$. As earlier, the values of the
energy parameters of the model in the units of $|t|$
\begin{eqnarray}\label{Data}
T=0.01,~t=-1,~t'=-0.65,~t''=-0.4,
\end{eqnarray}
were chosen such that the Fermi surface would have the form of the
pockets. The dashed line corresponds to the lower branch
$E^{-}_{\textbf{k}}$ of the spectrum; the solid line, to the upper
branch $E^{+}_{\textbf{k}}$. Despite there are two solutions of
the dispersion equation, in reality the only branch is revealed.
This is due to the fact that at $\Omega_{SF}\chi_0\rightarrow 0$
the partial contributions
\begin{eqnarray}
\label{Par_SP_Int} A^{\pm}_{\textbf{k}}=\frac{1}{2}C_n\left\{1 \pm
\textrm{sign}(t_{1\textbf{k}})\right\}.
\end{eqnarray}
sharply grow from zero to the end value or sharply drop from the
end value to zero at the variation in the quasimomentum (dashed
lines in Fig.~\ref{A_k}). Therefore, at each nonzero point of the
Brillouin zone having the same value, only one of the partial
spectral intensities $A^{\pm}(\textbf{k},\omega)$ may exist. The
total $A(\textbf{k},\omega)$ spectral intensity (the values of
these functions are reflected in the plots by thickness of the
corresponding lines) remains invariable at the variation in the
quasimomentum along the energy spectrum. In our simple case, this
corresponds to the Hubbard-I approximation. The considered
features are illustrated by the left plots of the second, third,
and fourth panels in Fig.~\ref{fig_spectra_SF}.

The situation becomes qualitatively different when SFs switch
(right plots in Fig.~\ref{fig_spectra_SF}). The upper right plot
presents the spectra calculated at $\Omega_{SF}\chi_0 = 0.16$. As
before, the solid line shows the $E^+_{\textbf{k}}$ spectrum; the
dashed line, the spectrum $E^-_{\textbf{k}}$. Switching of the
interaction between the electron subsystem with the subsystem of
spin degrees of freedom via the kinematic mechanism leads to
repulsion of the spectrum branches at the points where they were
osculating without interaction.

The more significant effect of SFs is related to considerable
modification of the spectral weights $A^{\pm}_{\textbf{k}}$, which
manifests itself in strong renormalization of the values of these
functions (Fig.~\ref{A_k}). As a result, the partial spectral
intensity $A^{+}(\textbf{k},\omega)$ acquires a finite value over
the entire curve of the energy spectrum $E^+_{\textbf{k}}$
(similarly, $A^{-}(\textbf{k},\omega)$ is finite along the entire
$E^-_{\textbf{k}}$ curve) and becomes strongly modulated. It can
be clearly seen from comparison of the left and right plots in the
second and third panels in Fig.~\ref{fig_spectra_SF}. The total
spectral intensity $A(\textbf{k},\omega)$ acquires the structure
reflecting the presence of two branches exhibiting strong
modulation. In the right plot of the lower panel in
Fig.~\ref{fig_spectra_SF}, the degree of darkening of the parts in
the vicinity of the points $(\textbf{k},\omega)$ corresponds to
the value of $A(\textbf{k}, \omega)$ in these points. It can be
seen that the allowance for SFs has led to induction of the shaded
zone and redistribution of the spectral intensity between the
basic and shaded zones.

The feature related to the change in the sign of parameter $t$ of
hopping to the first coordination sphere (Fig.\ref{SI_FS_SF}) is
characteristic of this model. Fig.~\ref{SI_FS_SF2} shows the
spectral intensity calculated along the principle directions of
the Brillouin zone (left plots) and at the Fermi contour for a
quarter of the Brillouin zone (right plots) at different signs of
$t$. It can be seen that, at $t>0$, $A(\textbf{k},\omega)$ is
maximum in the part of the Fermi contour that is closer to the
point $(0,0)$, which is consistent with the experimental data.

\section{\label{sec:level8}Summary}

To sum up, we formulate the principles of the kinematic formation
of modulation of the spectral intensity $A(\textbf{k},\omega)$. Of
fundamental importance is the use of the faithful representation
for a single-fermion Matsubara Green's function
$D(\textbf{k},i\omega_m)$. For the kinematic mechanism,
$D(\textbf{k},i\omega_m)$ is expressed as the product of the
propagator part and strength operator $P(\textbf{k}, i\omega_m)$. The
presence of the strength operator in the numerator of the Green's
function and its dependence on the Matsubara frequency and
quasimomentum lead to the fact that the isoenergetic lines in the
quasimomentum space become different from the lines where the
strength operator has a constant value. This spacing is one of the
causes of modulation of the spectral intensity
$A(\textbf{k},\omega)$. It is important that the integral of
hopping to the first coordination sphere determines the Fermi
contour part where $A(\textbf{k},\omega)$ considerably decreases.

The specific cause of $A(\textbf{k},\omega)$ modulation in the
framework of the kinematic interaction is that SFs lead to the
formation of a shaded zone, which represents the initial zone
shifted in the quasimomentum space by the vector $(\pi,\pi)$. As a
result, the total pattern of the spectrum is formed by coherent
hybridization of these two zones, between which the spectral
intensity is redistributed and, consequently, the density of
states in the vicinity of the chemical potential drops.

To demonstrate the kinematic formation of the pseudogap behavior
as brightly as possible, we considered the Hubbard model in the
limit of strong correlations when the dynamics of Hubbard fermions
is governed by the kinematic interaction. Obviously, the
discovered mechanism of the pseudogap phase formation is universal
and relevant for other models of strongly correlated systems.

\begin{acknowledgments}
We thank R.\,O. Zaitsev and N.\,M. Plakida for their criticism and
useful discussions of the results.

The study was supported by the program "Quantum physics of
condensed matter" of the Presidium of the Russian Academy of
Sciences (RAS); the Russian Foundation for Basic Research (project
No. 10-02-00251); the Siberian Branch (SB) of RAS
(Interdisciplinary Integration project No. 53); Federal
goal-oriented program "Scientific and Pedagogical Personnel for
Innovative Russia 2009-2013". Two of authors (A. G. and M. K.)
would like to acknowledge the support of Grant of President of
Russian Federation (project No.~MK-1300.2011.2) and Lavrent'ev's
Competition of Young Scientists of SB RAS.
\end{acknowledgments}

\end{document}